# Coupling and decoupling between translational and rotational dynamics in a tetrahedral molecular liquid

Gan Ren (任淦)[†]

School of Science, Civil Aviation Flight University of China, Guanghan 618307, China

**Abstract**

The translational dynamics usually decouples earlier than the rotational in supercooled liquids as the temperature decreases, whereas the inverted scenario remains scarce reported. In this work, starting from the coarse-grained ortho-terphenyl model, we build a rigid tetrahedral structure model. It exhibits an earlier decoupling in rotation than in translation. The Stokes-Einstein-Debye relation breaks down while the Stokes-Einstein relation remains intact. The decoupling happens at approximately $2T_\mathrm{g}$ similar as that observed in supercooled water. The rotation shows more heterogeneous dynamics than the translation at all temperatures. Our results suggest that the steric hindrance plays an important role in determining the decoupling between the translational and rotational dynamics.

## 1. Introduction

Over the past three decades it has become clear that the dramatic slowing down dynamics of supercooled liquids is accompanied by a progressive loss of correlation between different kinds of molecular motion. One of the clearest manifestations is the decoupling of translational and rotational diffusion [1, 2]. In the high-temperature liquid state, molecular motion is largely homogeneous and follows well-established hydrodynamic laws. The Stokes-Einstein (SE) relation $D_t = k_\mathrm{B}T/C\eta a$ or $D_t \sim \tau_t^{-1}$ governs the translational diffusion $D_t$, and the Stokes-Einstein-Debye (SED) relation $D_r = k_\mathrm{B}T/C'\eta a^3$ or $D_r \sim \tau_{rn}^{-1}$ governs the rotational diffusion $D_r$, where $k_\mathrm{B}$ is the Boltzmann constant, $T$ the temperature, $\eta$ the bulk viscosity, $a$ the effective hydrodynamic radius, $\tau_t$ the structural relaxation time, $\tau_{rn}$ the rotational relaxation time, $C$ and $C^{'}$ are constants determined by the boundary conditions, where "~" means "proportional". It implies a fundamental coupling between these two diffusion constants. However, the dynamics of liquids become increasingly heterogeneous and non-exponential when undergoes supercooling towards the glass transition point $T_g$, and the two diffusion constants are decoupled from the macroscopic viscosity or relaxation time [3]. The decoupling implies SE relation or SED relation is breakdown, or both fails. By combination of the two formulas for SE relation and SED relation as well as the assumption of constant $a$, many different formulas were adopted to verify the decoupling, such as whether $D_t/D_r$, $D_t\tau_{rn}$ or $T\tau_{rn}/\eta$ keep as a constant as condition changes.

Fujara et al. [4] provided the first direct NMR evidence in supercooled ortho-terphenyl (OTP). They observed the $D_t$ is proportional to $\eta^{-1}$ above $1.2T_g$, and otherwise follows a fractional form like $D_t \sim \eta^{-0.75}$. However, the $D_r$ remains proportional to $\eta^{-1}$ down to $T_g$. Mapes et al. [5] measured the diffusion constant of OTP near $T_g$, and found the combination of their data and the data of Fujara can be well fitted by $D_t \sim \eta^{-0.8}$. Cicerone et al. [6, 7] found the $D_t \sim T/\eta$ and $D_t\tau_{rn}$ are size dependent for the probes in OTP, the SE relation described by $D_t \sim T/\eta$ fails with cooling, and the $D_t\tau_{rn}$ changes relatively small for large sized probe but increases almost two orders of magnitude towards $T_g$ for the small probe. Eastwood and coworkers [8] reported that decoupling sets in at approximately $1.2T_g$ based on $D_t\tau_{r2}$, with quantitative agreement between simulation and experiment. Stillinger and co-workers [9] observed the scaled ratio $D_t/D_r$ is almost equal to 1.0 within 260-346K, but it started to deviate when $T < 260$K. The scaled simulated and experimental $D_t/D_r$ are almost equal to 1.0 at high temperature but deviate from 1.0 at a certain temperature upon cooling.

The decoupling phenomena were also intensively investigated in supercooled water. The $D_t\tau_{rn}$ is observed not to be a constant but are temperature and density dependent in the supercooled SCP/E,

TIP5P and ST2 water [10-14]. The decoupling in $D_t\tau_{r2}$ happens at $2T_g$ in TIP5P and ST2, but not the usually recognized $1.2T_g$, which can be explained by a two-state scenario [15, 16]. The $D_r\tau_t/T$ is almost a constant in SPC/E water when $T$ > 280K but starts to increase as temperature decreases [17]. The ratio $D_t/D_r$ is not a constant but decreases with cooling in SPC/E[17] and TIP4P/2005 [18]. Both $D_r \sim T/\tau_t$ and $\tau_{r2}^{-1} \sim T/\tau_t$ get a crossover in ST2 water as temperature decreases [19]. The $\tau_{r2}^{-1} \sim T/\tau_t$ fails in the whole simulated temperature range; however, $D_r \sim T/\tau_t$ holds at high temperatures and otherwise fails in a fractional form.

The similar phenomena are also observed in colloid systems with an increasing volume fraction $\phi$. Weeks and co-workers [20] visualized supercooled colloidal fluids and showed that the $D_t$ began to violate the SE relation prediction at $\phi \approx 0.52$, whereas the $D_r$ did not deviate from the SED relation until $\phi \approx 0.56$–$0.57$. Subsequent work on quasi-2D colloidal fluids further nuanced this picture, showing that the degree of decoupling is sensitive to local particle geometry, with short dimers exhibiting a different decoupling pattern than long ones, thereby linking the phenomenon to local caging effects [21].

The studies cited above generally report that translational degree decouples earlier than rotational one. The prevailing theoretical interpretation attributes this asymmetric decoupling to the spatially heterogeneous and facilitated nature of dynamics in supercooled liquids. In this framework, translation is dominated by rare collective string-like motions that percolate through transient fast regions, allowing it to escape the average viscous drag more readily [22]. In contrast, rotational motion is thought to be more localized and sensitive to the relaxation of the immediate cage formed by neighboring particles, making it more closely to the average structural relaxation time and viscosity [23]. This interpretation is strongly supported by molecular dynamics simulations of simple model liquids, which have visualized these heterogeneous regions and connected the onset of decoupling to distinct stages in the system's exploration of its potential energy landscape [24].

However, this scenario is observed to be inverted in anisotropic ellipsoids and dumbbells. Quasi-2D colloidal ellipsoids with aspect ratio 6 showed that the ratio $D_t/D_r$ increased by more than one decade across the glass transition, and a lower density of glass transition in rotational degree is observed than the translational, which implies the rotation decouples earlier than the translation [25, 26]. Molecular dynamics simulations of elongated rods reproduce the same trend when the rotational cooperatively length grows faster than the translational one [27]. A dense mixed dumbbells liquid also shows an earlier decoupling of the rotational degree than the translational one at $1.2T_g$ by comparing $D_t \sim \tau_t^{-1}$ and $D_r \sim \tau_t^{-1}$ [22].

These special cases underscore that the order of decoupling is not immutable; rather, it is governed by the relative growth of translational versus rotational dynamic length scales. While the existing experimental or simulated realizations are still sparse, and only exist in anisotropic ellipsoids and dumbbells, a further search for additional model systems is warranted. In this work, we introduce a rigid tetrahedral molecular model as a complementary system to study translation-rotation decoupling. Unlike conventional models such as OTP where translational diffusion decouples prior to rotational relaxation, our tetrahedral system exhibits an inverse decoupling sequence: rotational dynamics decouple at approximately $2T_g$ while translational diffusion remains coupled to viscosity and relaxation time to lower temperatures. This inversion originates from the unique isotropic steric hindrance of the tetrahedral geometry, which suppresses rotation more effectively than translation. To our knowledge, this represents the first observation of rotation-preceded decoupling in a tetrahedral molecular glass former.

## 2. Simulation details and analysis methods

The configuration for the molecular dynamics (MD) simulation is consisted of 2048 rigid tetrahedral molecules (TM) in a cubic box with side length 9.676 nm. Using the Lennard-Jones potential diameter, and accounting for both the tetrahedral volume and the molecular volume outside the tetrahedron, the volume fraction is approximately $\phi \approx 0.54$ . The TM model is based on the coarse-grained OTP model [28, 29]. It is consisted with four coarse-grained atoms in OTP with the same bond length and interaction parameters as OTP. All our MD simulations were carried out with the GROMACS package [30, 31]. The periodic boundary conditions were applied in all three directions of the Cartesian space. The van der Waals interactions were calculated with a cutoff of 1.4 nm. Thirty four temperatures were simulated and is distributed within 420-1600K. The temperature was kept at a constant by the Nosé-Hoover thermostat [32, 33].

The $D_t$ is calculated via the mean square displacement as

$$D_t = \lim_{t\to\infty} \sum_{i=1}^{N} \left\langle \left| \vec{r}_i(t) - \vec{r}_i(0) \right|^2 \right\rangle \Big/ 6Nt \,, \qquad (1)$$

where $\vec{r}_i(t)$ is the center of mass of $i$th TM at time $t$, < > denotes a time average, $N$ is the number of TM. The $\tau_t$ is determined by the self-intermediate scattering function [34]

$$F_s(k,t) = \frac{1}{N} \sum_{j=1}^{N} \left\langle \mathrm{e}^{\mathrm{i}k\cdot\left[\vec{\mathbf{r}}_j(t) - \vec{\mathbf{r}}_j(0)\right]} \right\rangle, \qquad (2)$$

where $N$ is the number of molecules, wavevector $k = 9.0\,\mathrm{nm}^{-1}$ corresponding to the first maximum of the static structure factor. And $\tau_t$ is determined by $F_s(k,\tau_t) = \mathrm{e}^{-1}$. The $D_r$ is calculated via its asymptotic relation with the rotational mean square displacement [17, 18]

$$D_r = \lim_{\Delta t\to\infty} \sum_{i=1}^{N} \left| \vec{\varphi}_i(t+\Delta t) - \vec{\varphi}_i(t) \right|^2 \Big/ 4N\Delta t \,, \qquad (3)$$

where $\vec{\varphi}_i(\Delta t)$ is the angular displacement. The rotational correlation time $\tau_{rn}$ is calculated via the rotational correlation function [18, 35]

$$C_n(t) = \sum_{i=1}^{N} \left\langle P_n\left[ \vec{e}_i(t) \cdot \vec{e}_i(0) \right] \right\rangle \Big/ N \,, \qquad (4)$$

where $P_n(x)$ is the $n$-th order Legendre polynomial, $\vec{e}_i(t)$ is the unit vector of $\vec{\varphi}_i(t)$, and $\tau_{rn}$ is determined by $C_n(\tau_{rn}) = \mathrm{e}^{-1}$.

The $\eta$ is determined by the method proposed by Hess [36] for its reliability and fast convergence in the linear response regime. Since the SE relation is a combination of Einstein relation $D_t = k_\mathrm{B}T/\alpha$ and Stokes's law $\alpha = C\eta a$. The $\alpha$ is determined by introducing a small force $f_e$ on a part of particles in the linear response regime, and 128 TMs are chosen in our simulation. The translational dynamic heterogeneity is usually described by the translational non-Gaussian parameter [37] as

$$\alpha_{2t}(t) = 3\langle r^4(t)\rangle / 5\langle r^2(t)\rangle^2 - 1. \quad (5)$$

And the translational dynamic heterogeneity is characterized by the rotational non-Gaussian parameter [38] as

$$\alpha_{2r}(t) = 3\left\langle \vec{\varphi}^4(t)\right\rangle \Big/ 5\left\langle \vec{\varphi}^2(t)\right\rangle^2 - 1. \quad (6)$$

**3. Results and discussion**

The calculated translational diffusion coefficient $D_t$, rotational diffusion coefficient $D_r$, viscosity $\eta$, structural relaxation time $\tau_t$, rotational relaxation time $\tau_{rn}$ for $n$ = 1-6 and frictional coefficient $\alpha$ (scaled by the mass of molecule TM) under different temperature $T$ are plotted in Fig. 1. The data are usually proposed to follow a Arrhenius law or a VFT law [34]. The fittings show the data can be well described by a VFT law $A = A_0 \mathrm{e}^{\pm E_a/(T-T_0)}$ other than the $D_r$, which can be well fitted by a Arrhenius law. In practice, we fitted the logarithm of the data to the VFT law in the form $\ln A = \ln A_0 \pm E_a/(T-T_0)$. The fitted logarithm of prefactor $\ln A_0$, activation energy $E_a$ and the Vogel temperature $T_0$ are plotted in Fig. 1 and the fitted data for $\tau_{rn}$ are listed in Table 1. The fittings show the system behaves more likely as a fragile liquid than a strong one [39]. The trend of $D_t$, $\eta$ and $\tau_{r2}$ versus $T$ are similar as that observed in previous experiments for OTP [4, 5, 40]. McCall et al. [40] show that their $D_t$ directly follows the VFT law with $E_a$ = 689 and $T_0$ = 231K. Their $T_0$ is close to our value of 220.6K; however our $E_a$ = 1573.89 is much greater than theirs. The results suggest that the rigid tetrahedral structure either makes the molecular hopping more difficult or requires more free volume for diffusion, consistent with the free volume theory [34]. The $T_g$ is conventionally defined in the literature as the temperature at which the viscosity reaches $10^{12}$ Pa·s [1]. Our VFT fit yields a value of $T_g \approx$ 314.5K. It is greater than the $T_g$ = 243K for OTP observed in experiments [5], suggesting that the rigid tetrahedral structure makes the system more prone to glass formation.

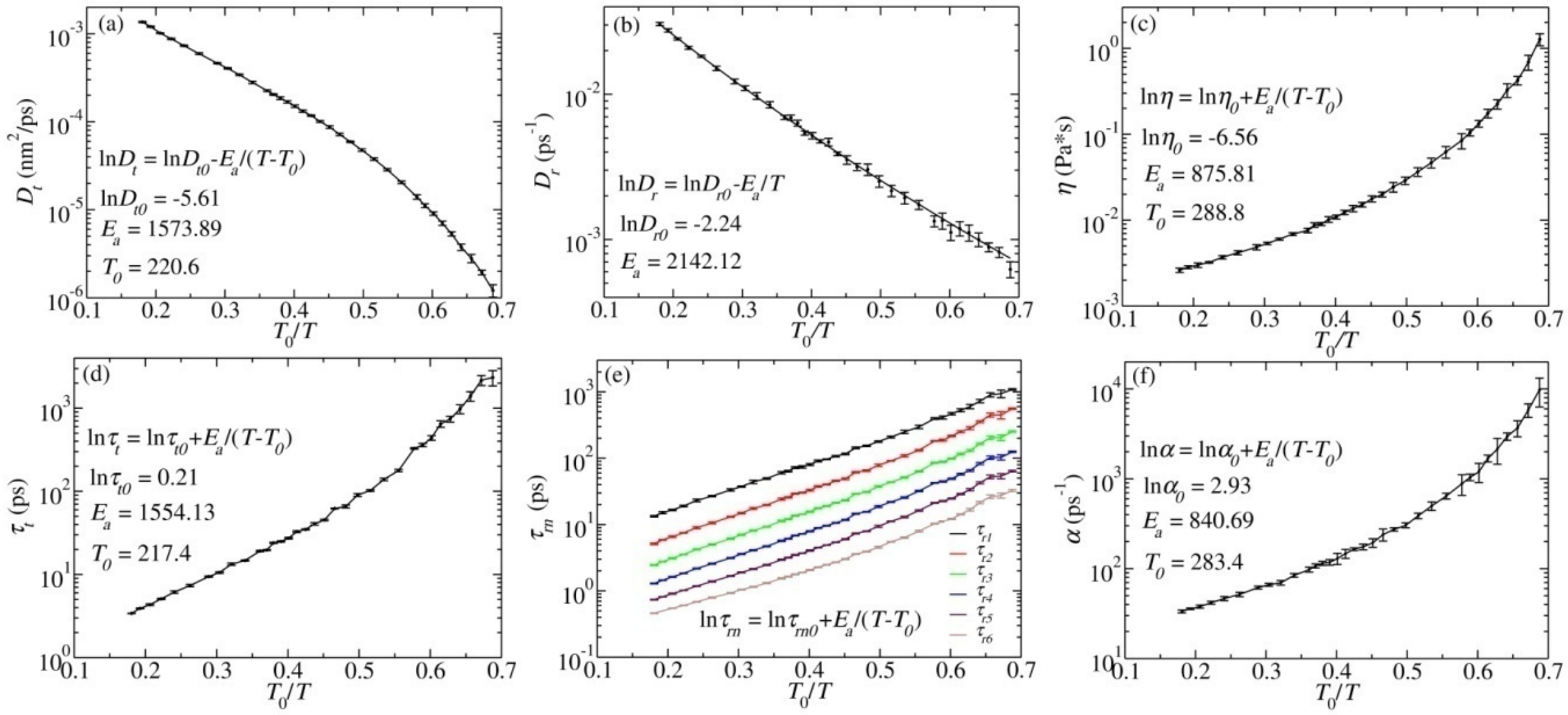

**Figure 1.** The $D_t$ , $D_r$ , $\eta$ , $\tau_t$ , $\tau_{rn}$ for $n$ = 1-6 and $\alpha$ as a function of $T_0/T$. (a) $D_t$ vs $T_0/T$; (b) $D_r$ vs $T_0/T$ ; (c) $\eta$ vs $T_0/T$; (d) $\tau_t$ vs $T_0/T$ ; (e) $\tau_{rn}$ vs $T_0/T$ ; (f) $\alpha$ vs $T_0/T$. $T_0$ = 288.8K and is obtained by fitting the $\eta$ with VFT law.

**Table 1.** The fitted logarithm of prefactor $\ln\tau_{rn0}$, activation energy $E_a$ and the Vogel temperature $T_0$ for $\tau_{rn}$ with $n$ = 1-6.

| | $\tau_{r1}$ | $\tau_{r2}$ | $\tau_{r3}$ | $\tau_{r4}$ | $\tau_{r5}$ | $\tau_{r6}$ |
|---|---|---|---|---|---|---|
| $\ln\tau_{rn0}$ | 1.00 | 0.43 | -0.35 | -0.95 | -1.41 | -1.68 |
| $E$ | 2475.9 | 1918.2 | 1960.5 | 1884.8 | 1684.0 | 1338.7 |
| $T_0$ | 0.162 | 94.20 | 86.71 | 94.80 | 119.3 | 163.6 |

To examine whether the translational and the rotational dynamics decouple, we actually need to verify the validity of SE relation and SED relation. Since the SE relation is actually originated from the combination of Einstein relation $D_t = k_B T/\alpha$ and Stokes law $\alpha = C\eta a$. If assuming $a$ to be a constant, the SE relation can be described by $D_t \sim T/\eta$. And the relation $D_t \sim \tau_t^{-1}$ holds exactly when dynamic heterogeneity is absent. So we adopt the two formulas $D_t \sim (T/\eta)^{-\xi_1}$ and $D_t \sim \tau_t^{-\xi_2}$ to test the SE relation, and adopt $D_t \sim (T/\alpha)^{-\xi_3}$ to test the original Einstein relation. The original SED relation proposed by Debye [35] is $D_r \sim \tau_{rn}^{-1}$ and $D_r = k_B T/\varsigma$, the latter can be expressed as $D_r = k_B T / C'\eta a^3$ with the Stokes formula $\varsigma = C'\eta a^3$, and is further expressed as $D_r \sim T/\eta$ with assuming $a$ is a constant. The $a$ is also evaluated by $a \sim \alpha/\eta$ by considering its potential changes. Therefore the SED relation is tested by $D_r \sim (T/\eta)^{\xi_4}$, $D_r \sim \tau_{rn}^{-\xi_{5n}}$ and $D_r\alpha^3 \sim (T\eta^2)^{\xi_6}$. Because the standard errors shown in Fig. 1 are smaller than 0.1 from 1600 K down to 450 K, and only three data groups within 420-440K exhibiting slightly larger uncertainties of 0.1–0.2. The fitting is mainly determined by the data in the range 450-1600 K, so if the corresponding exponent $\xi_i = 1.0$ ($i$=1-6) lies within 0.9-1.1, the SE or SED relation is satisfied within statistical error; otherwise, it is invalid.

As the Fig. 2 shown, the SE relation $D_t \sim T/\eta$, $D_t \sim \tau_t^{-1}$ and $D_t \sim T/\alpha$ follow fractional forms as $D_t \sim (T/\eta)^{\xi_1}$, $D_t \sim \tau_t^{-\xi_2}$ and $D_t \sim (T/\alpha)^{\xi_3}$; and the exponents are $\xi_1 = 0.956$, $\xi_2 = 1.029$ and $\xi_3 = 1.023$, respectively. All exponents are so close to the exact result $\xi_i = 1.0$, which indicates the SE relation is valid in the simulated temperature range. The $D_t \sim \tau_t^{-1}$ is an exact result when the translation displacement $\Delta\vec{r}_i(t)$ follows Gaussian distribution. However, Kawasaki and Kim have shown that the validity of $D_t \sim \tau_t^{-1}$ depends on the ratio of the cage-breaking time scale to structural relaxation [41]. The cage-breaking time is characterized by $t_{\max}$, defined as the time corresponding to the maximum of the $\alpha_{2t}(t)$. When the ratio $t_{\max}/\tau_t$ is not much smaller than unity (e.g., ~0.1), diffusion and relaxation remain coupled and the $D_t \sim \tau_t^{-1}$ is preserved. As shown in Fig. 3(c), the smallest ratio at $T$ = 420K is so close to 0.1. This explains why, despite the maximum $\alpha_{2t}(t)$ being around 1.5 at $T$ = 420K as shown in Fig. 3(a), the $D_t \sim \tau_t^{-1}$ remains valid. The discrepancies in exponents between $D_t \sim (T/\eta)^{\xi_1}$ and $D_t \sim (T/\alpha)^{\xi_3}$ are due to the assumption of a constant *a* adopted in $D_t \sim T/\eta$. By comparing $D_t \sim (T/\eta)^{\xi_1}$ and $D_t \sim (T/\alpha)^{\xi_3}$, the *a* should be varied with $D_t$ like $a \sim D_t^{0.0685}$ and decreases with decreasing temperature. In case of the small differences between $\xi_1$ and $\xi_3$, the temperature dependence is very small. Nevertheless, these modest variations still induce discrepancies of approximately 0.08 between the exponent $\xi_1$ and $\xi_3$, and which propagate into much larger differences in subsequent test of the SED relation. This suggests that the exponent *a* serves as a sensitive diagnostic for the validity of the SE relation. Such behavior aligns with our earlier observation that *a* exhibits a downward trend upon cooling in systems dominated by attractive interactions [42].

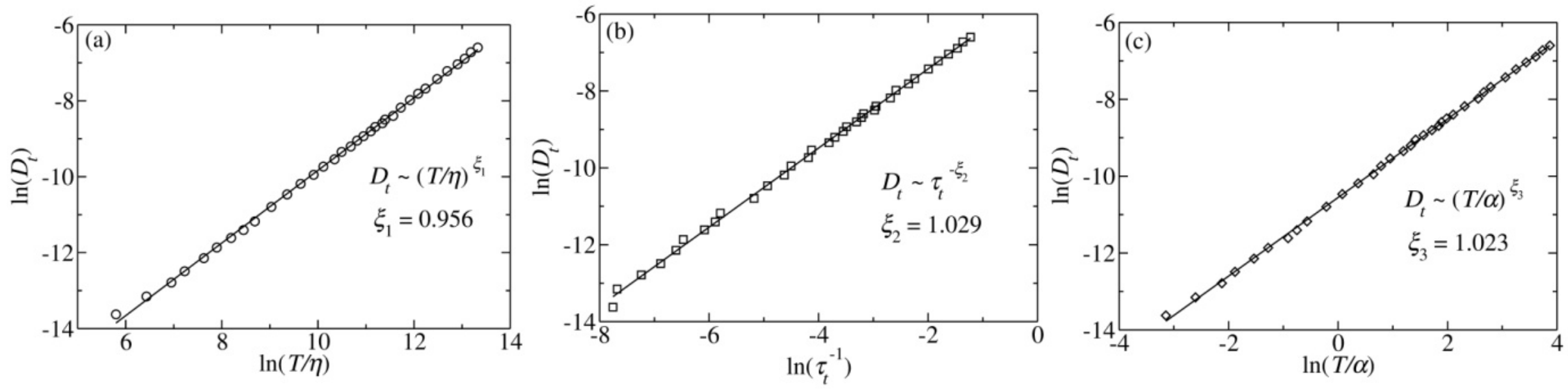

**Figure 2.** Verification of the validities of the SE relation: (a) $D_t \sim T/\eta$; (b) $D_t \sim \tau_t^{-1}$; (c) $D_t \sim T/\alpha$. The calculated data are represented by circles and solid lines are fitting with $D_t \sim (T/\eta)^{\xi_1}$, $D_t \sim \tau_t^{-\xi_2}$ and $D_t \sim (T/\alpha)^{\xi_3}$, respectively.

The simulated results of SED relation $D_r \sim T/\eta$ plotted in Fig. 4(a) are not fallen on a line but with a crossover at temperature $T_x \approx 600\text{K}$. The exponents for the two parts are $\xi_4 = 0.818$ and 0.289, respectively. Both fitted exponents depart from the theoretical value of $\xi_4 = 1.0$, lying beyond the validity interval 0.9-1.1. Notably, the deviation is significantly attenuated in the high-temperature regime $T > T_x$ compared with the low-temperature regime $T < T_x$. The result indicates the breakdown of the SED relation described by $D_r \sim T/\eta$ in the whole simulated temperature range.

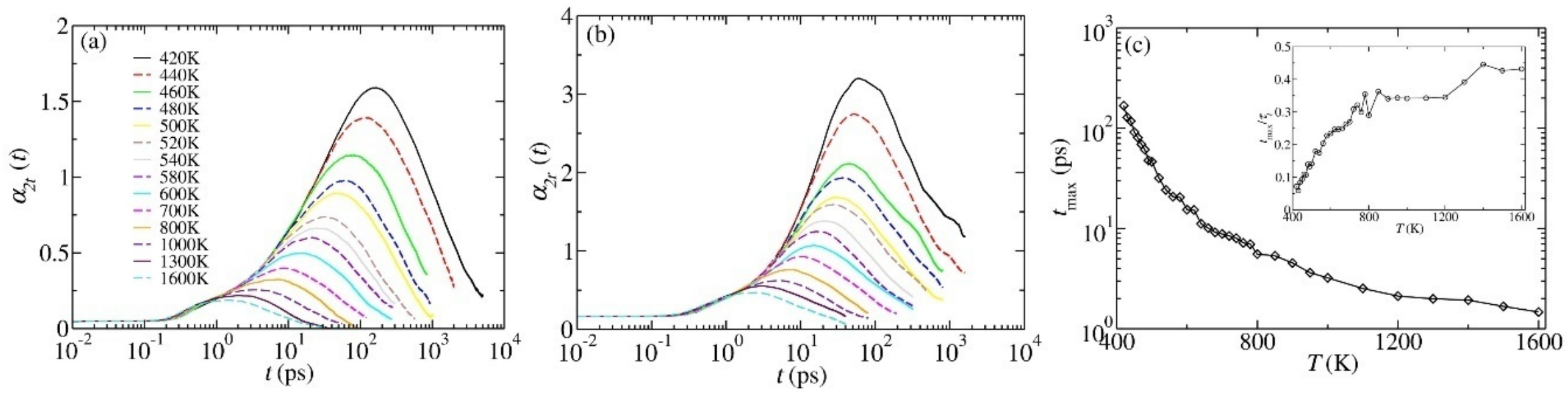

**Figure 3.** (a)The translational non-Gaussian parameter $\alpha_{2t}(t)$; (b) the rotational non-Gaussian parameter $\alpha_{2r}(t)$; (c) the $t_{\max}$ corresponding to the maximum of $\alpha_{2t}(t)$ as a function of $T$, the inset is the ratio of $t_{\max}$ to $\tau_t$.

A similar crossover is also observed around 600K in the SED relation described by $D_r \sim \tau_{rn}^{-1}$ for $n$ = 1-6, but with different exponents comparing with $D_r \sim T/\eta$. As the fitted $\xi_{5n}$ listed in Table 2 shown, the exponents in $D_r \sim \tau_{rn}^{-\xi_{5n}}$ for $n$ = 1-5 is so close to the exact result $\xi_{5n} = 1.0$ above 600K, which indicate the validity of $D_r \sim \tau_{rn}^{-1}$. However, $\xi_{5n} \simeq 0.6$ implies the breakdown of $D_r \sim \tau_{rn}^{-1}$ for $n$ = 1-5 below 600K. Moreover, the $D_r \sim \tau_{r6}^{-1}$ is breakdown in the whole temperature range and with $\xi_{56} = 1.227$ for $T$ > 600K and otherwise 0.585. The result is consistent with $\alpha_{2r}(t)$ plotted in Fig. 3(b), which is close to Gaussian distribution at higher temperatures but deviates as temperature decreases. The results are similar with observations in TIP4P/2005, where the $D_r \sim \tau_{rn}^{-1}$ is more strongly violated at larger $n$ [18]. As point out in ref. [18] , lower order $n$ probe large angle molecular reorientation, while high-order probes small-angle motion within the cage, the two are fundamentally different physical processes with distinct time scales. At high temperatures, molecules rotate freely, cage effects are minimal, and both processes remain coupled to thermal motion. Lower order $\tau_{r1-5}$ follows Debye relaxation, matching the $D_r$, so $D_r \sim \tau_{r1-5}^{-1}$ holds. At low temperatures, dynamic heterogeneity emerges. Cage escape becomes rare and heterogeneous, slowing $\tau_{r1-5}$ dramatically. However, $n$ = 6 still probes fast in-cage wobbling. Thus $\tau_{r1-5}$ and $D_r$ decouple, $D_r \sim \tau_{r1-5}^{-1}$ breaks down, while $\tau_{r6}$ was never synchronized with $D_r$ due to this inherent time-scale separation.

The $\alpha_{2r}(t)$ is larger than $\alpha_{2t}(t)$ at any temperature, which indicates the system has a more heterogeneous rotational dynamics than the translational dynamics at any temperature. This arises from steric hindrance, as observed in colloidal ellipsoids [25, 26] and elongated rods [27]. The rigid tetrahedral structure and high volume fraction hinder molecular rotation more severely than translation, since steric effects require neighboring molecules to rotate collectively to accommodate the central molecule's reorientation.

The data for the SED relation $D_r = k_B T / C^{'} \eta a^3$ tested by $D_r \alpha^3 \sim (T\eta^2)^{\xi_6}$ plotted in Fig. 4(c) is also shown a crossover around $T_x \approx 600\text{K}$. The fitted exponent is $\xi_6 = 1.051$ above $T_x$ and otherwise $\xi_6 = 1.262$. The former is close to the exact $\xi_6 = 1.0$ and indicates the $D_r = k_B T / C^{'} \eta a^3$ is established when $T > T_x$. The latter lies beyond the validity interval and implies the breakdown for $T < T_x$. Comparing the results given by $D_r \sim T/\eta$ and $D_r = k_B T / C^{'} \eta a^3$, the results given by $D_r \sim T/\eta$ takes some corrections after considering the variation of $a$ into account. The $D_r \sim T/\eta$ is almost corrected to the exact result $\xi_6 = 1.0$ for $T > T_x$. However, the $\xi_4 = 0.289$ for $T < T_x$ is corrected to $\xi_6 = 1.262$. Although the correction make the breakdown become smaller, yet the $D_r = k_B T / C^{'} \eta a^3$ is still invalid for $T < T_x$. The result also signify the temperature dependent of $a$ and its significance in testing of the validity SED relation.

Combined the results given by Fig. 2 and Fig. 4, the SE relation are all valid for three forms but the three formulas of SED relation are all breakdown. The decoupling is existed in the whole simulated temperature range by comparing the three SE relations with the SED relation described by $D_r \sim T/\eta$ and $D_r \sim \tau_{r6}^{-1}$. Comparing the results given by the three SE relations with the SED relation described by $D_r \sim \tau_{rn}^{-1}$ for $n$=1-5 and $D_r = k_B T / C^{'} \eta a^3$, the translational and rotational motion are still coupling for $T > T_x$ but the two decouples when $T < T_x$. The $T_x$ and $T_g$ nearly satisfy the relation $T_x \approx 2T_g$, which is similar as that observed in TIP5P and ST2 water [13]. However, it is different from that observed in OTP [4, 9] and dumbbell molecular systems [22], which shows a decoupling at $1.2T_g$. And our systems show a similar earlier rotational decoupling than the translational motion as that observed in dumbbell molecular systems, but inverted in OTP [4, 9], TIP5P and ST2 water [13].

The phenomena observed can be explained by the differences in the mechanism of diffusion between the translational and the rotational [22, 23]. Translational diffusion is nonlocal, requiring escape from the cage trap. In contrast, rotational motion is more localized and sensitive to the relaxation of the immediate cage formed by neighboring particles. As observed by Kob and coworkers [22, 43] for a two-component dumbbell mixture at $\phi = 0.708$, rotational decoupling occurs earlier in this system despite its weak steric hindrance arising from the dumbbell geometry. This system resembles commonly simulated water or OTP, differing primarily in its significantly higher volume fraction. When steric hindrance and dense packing are present, rotation is typically more constrained than translation, leading to earlier decoupling of rotational motion. Our system is with a large volume faction $\phi = 0.54$, and the tetrahedral structure introduces strong steric hindrance to rotation. We can make a simple estimate of the steric hindrance for rotation. Assuming rotation about the center of mass, both the tetrahedral side length and the Lennard-Jones parameter $\sigma$ are equal to 0.483 nm, each molecule requires a circumscribed sphere of a tetrahedron of $0.65 \text{nm}^3$ to rotation freely, and the total volume need is approximately 1.47 times the simulation box volume. In comparison, for translation the molecule has an additional 0.46 fraction of free volume available for motion. Our rough estimate shows that steric hindrance has a stronger effect on rotation than on translation, leading to earlier rotational decoupling.

To further explore the possible influence of steric hindrance introduced dynamic heterogeneity on the translational and rotational dynamics, we calculated the $D_t$, $D_r$, $\tau_t$ and $\tau_{r3}$ for the fastest 7% and slowest 7% of molecules, following the approach in Ref. [17], and then examined the corresponding SE relation $D_t \sim \tau_t^{-1}$ and SED relation $D_r \sim \tau_{r3}^{-1}$, the correlated data are plotted in Fig .5. The $D_t$ and $\tau_t$ for the fastest and slowest subsets can differ by up to 50 fold, indicating strongly heterogeneous translational dynamics. The $D_t \sim \tau_t^{-1}$ holds for the slowest 7% of molecules as shown by $\xi_s = 1.004$ but breaks down for the fastest with $\xi_f = 1.93$. While it has been argued that $D_t$ is determined by mobile molecules whereas $\tau_t$ is governed by immobile ones [34], we examined a mixed SE relation $D_t \sim \tau_t^{-1}$ using the $D_t$ of the fastest and the $\tau_t$ of the slowest. Despite the large discrepancy between these two subsets, the SE relation remains valid in this mixed case as shown by the green line in Fig. 5(a), consistent with the results shown in Fig. 2.

Compared with the $D_t$ and $\tau_t$ for the fastest and slowest subsets, no such large difference is observed in $D_r$ or $\tau_{r3}$, and these differ by only up to 1.8 fold, suggesting that the rotational dynamics are more homogeneous than the translational ones. However, the SED relation $D_r \sim \tau_{r3}^{-1}$ for both the fastest and slowest subsets is only valid at high temperatures with $\xi_f = 0.925$ and $\xi_s = 1.02$, and breaks down at low temperatures as shown by $\xi_f = 0.853$ and $\xi_s = 0.854$, consistent qualitatively with the results shown in Fig. 4. The mixed SED relation $D_r \sim \tau_{r3}^{-1}$ using the $D_r$ of the fastest and the $\tau_{r3}$ of the slowest fails in the whole simulated temperature range with $\xi_{fs} = 0.871$. Given that the validity of the $D_t \sim \tau_t^{-1}$ and $D_r \sim \tau_{r3}^{-1}$ relies on Gaussian behavior, steric hindrance causes greater deviation from Gaussian dynamics in rotation than in translation, consistent with the observed non-Gaussian parameter.

The $\phi = 0.54$ is close to the decoupling displayed in spherical colloid system [20]. However, due to the rigid tetrahedral structure discussed above, each molecule requires a large spherical volume to rotate freely. In contrast, the decoupling in anisotropic particles such as dumbbells, ellipsoids, and elongated rods is due to due to the intrinsic translation-rotation coupling, prolate molecules can translate along their long axis via rotation, leading to a shape-dependent decoupling that varies with aspect ratio and packing fraction. Thereby it requires less free volume for rotational motion than our system. Consequently, these systems need a higher volume fraction to achieve earlier rotational decoupling relative to translation. By comparison, the effective volume fraction in our system is significantly lower than that in dumbbell systems [22, 43], colloidal ellipsoids [25, 26] and elongated rods [27], where the volume fraction $\phi > 0.7$. The steric hindrance with rigid tetrahedral structure is also reasons for the decoupling exhibiting temperature differences between our systems and the dumbbell system [22, 43]. Meanwhile, the decoupling temperature of approximately $2T_g$ is close to that observed in TIP5P and ST2 water models [13, 15, 16], which might suggest that the decoupling in TM is similarly related to its tetrahedral molecular structure. However, the underlying mechanisms are fundamentally different. In supercooled water, hydrogen bonding drives molecules into tetrahedral configurations upon cooling, triggering a disorder-to-order transition between two distinct liquid states. The emergence of a system-spanning hydrogen-bonded network is captured by the tetrahedral order parameter $Q_4$, which exhibits a pronounced increase as the transition proceeds. By contrast, our system features only isolated tetrahedral molecules.

To uncover the hidden resemblance, we calculated the local $Q_4$ based on the molecular center of mass [44]. The corresponding probability distribution and ensemble-average mean are present in Fig. 6. The probability distributions differ only slightly across the entire temperature range, from 1600K down to 420K. The mean $Q_4$ exhibits only a weak temperature dependence, increasing by merely 0.03 upon cooling from 1600 K to 420 K. Notably, no liquid–liquid phase transition is detected, in contrast to the behavior reported for supercooled water [44]. Moreover, previous studies have shown that the decoupling temperature is not universal but depends on the aspect ratio of rods [45], as well as the bond length of dimers and the volume fraction [21]. Therefore, we propose the similar decoupling temperatures observed in our system and supercooled water may be coincidental.

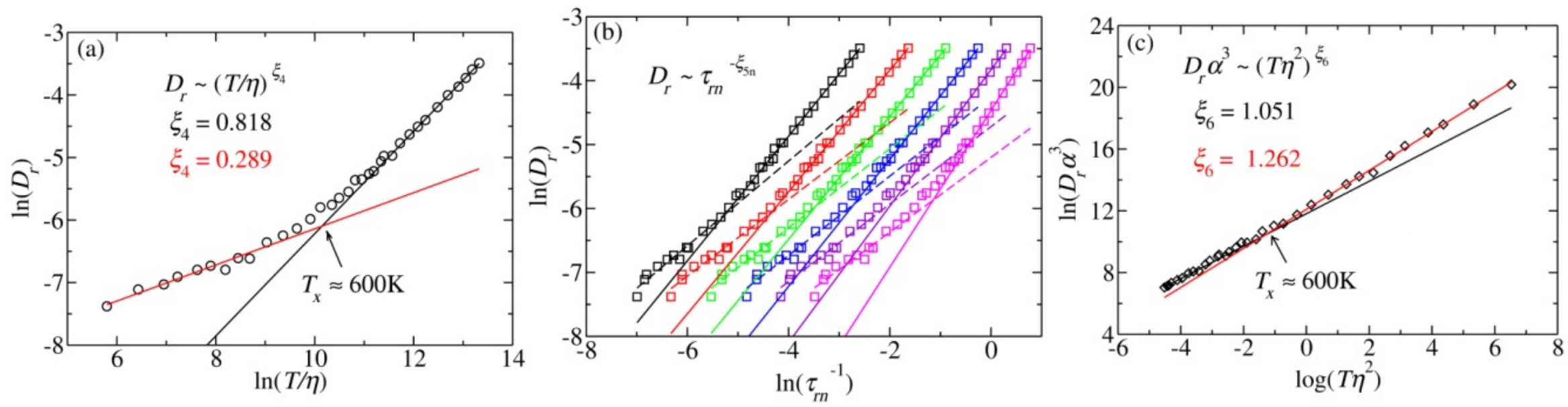

**Figure 4.** Verification of the validities of the SED relation: (a) $D_r \sim T/\eta$; (b) $D_r \sim \tau_{rn}^{-1}$; (c) $D_r \sim T/\eta a^3$.

The calculated data are represented by circles and fitted by $D_r \sim (T/\eta)^{\xi_4}$, $D_r \sim \tau_{rn}{}^{-\xi_{5n}}$ and $D_r\alpha^3 \sim (T\eta^2)^{\xi_6}$, respectively. The fitted exponent $\xi$ in (a) and (c) is written in the same color as the corresponding solid fitting line, and the data for (b) is listed in Table 2.

**Table 2.** The fitted exponent $\xi_{5n}$ for solid line and dotted line in $D_r \sim \tau_{rn}{}^{-\xi_{5n}}$.

| | $\xi_{51}$ | $\xi_{52}$ | $\xi_{53}$ | $\xi_{54}$ | $\xi_{55}$ | $\xi_{56}$ |
|---|---|---|---|---|---|---|
| solid | 0.663 | 0.600 | 0.619 | 0.620 | 0.606 | 0.585 |
| dotted | 0.974 | 0.952 | 0.962 | 0.994 | 1.066 | 1.227 |

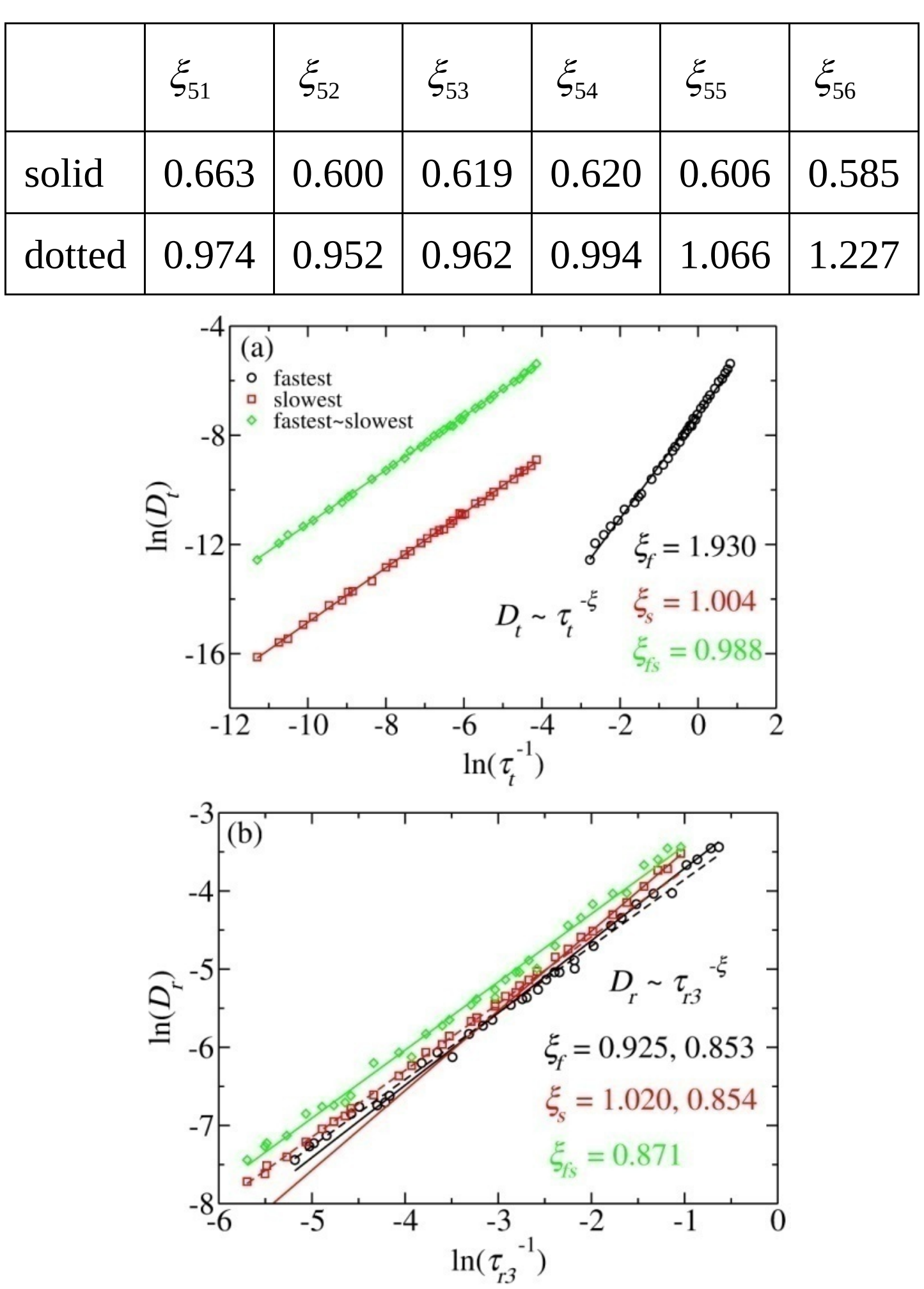

**Figure 5.** Verification of the validities of SE relation $D_t \sim \tau_t^{-1}$ and SED relation $D_r \sim \tau_{r3}^{-1}$ for the fastest 7% and slowest 7% of molecules. The green symbols represent data for the fastest 7% of molecules in terms of diffusion and the slowest 7% in terms of relaxation time.

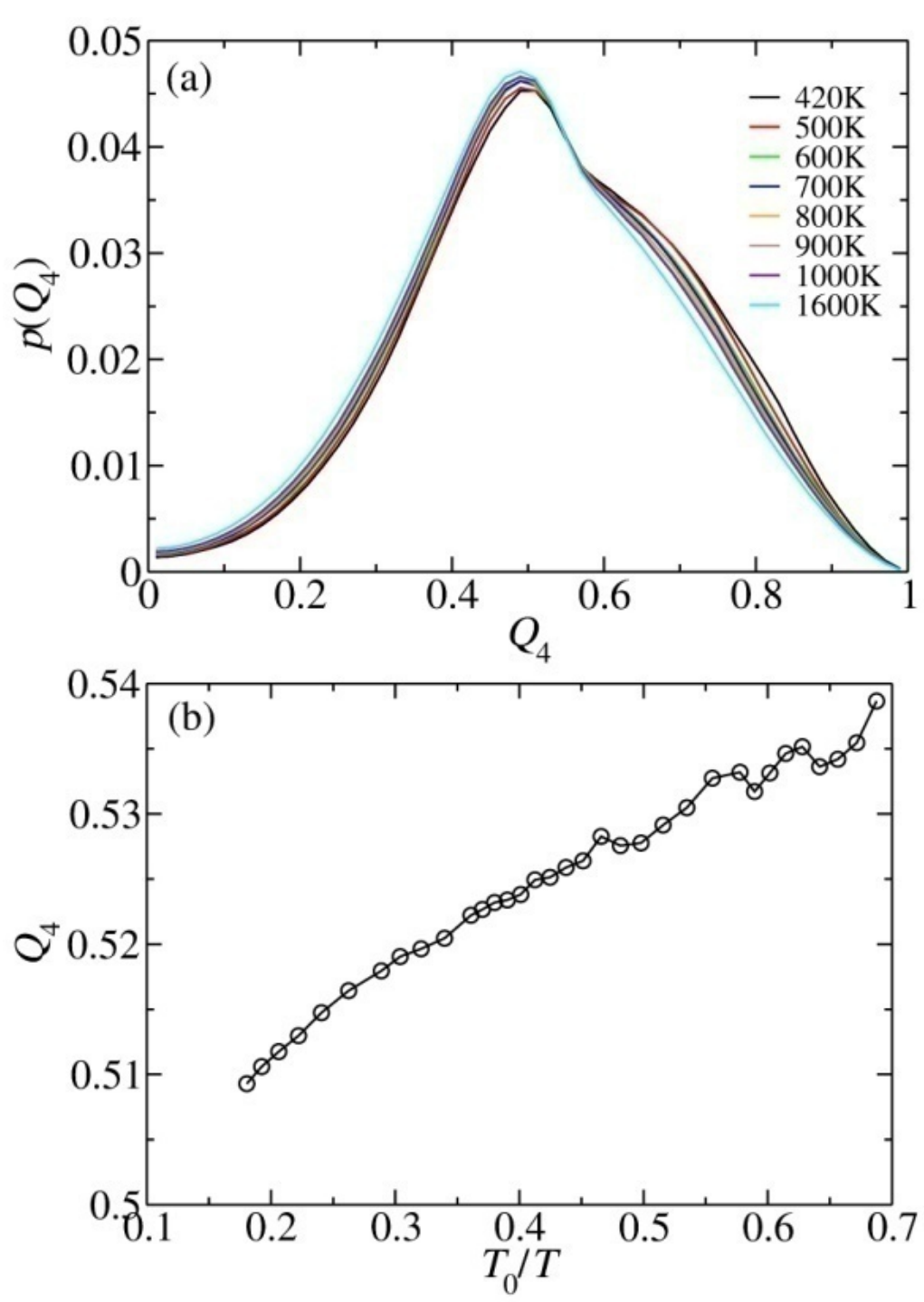

**Figure 6.** (a) Probability distribution of $Q_4$ at various temperatures; (b) temperature dependence of the mean $Q_4$.

## 4. Conclusions

In summary, we have examined the coupling and decoupling between the translational and rotational dynamics in a tetrahedral molecular liquid by testing the validity of the SE relation and SED relation. Our results indicate the SE relation described by $D_t \sim T/\eta$, $D_t \sim \tau_t^{-1}$ and $D_t \sim T/\alpha$ are all established. The SED relation described by $D_r \sim T/\eta$, $D_r \sim \tau_{rn}^{-1}$ and $D_r \sim T/\eta a^3$ all show a crossover around $T_x$ = 600K. The $D_r \sim T/\eta$ is breakdown in the whole simulated temperature range and follows a fractional form on both sides of $T_x$. The $D_r \sim \tau_{rn}^{-1}$ for n = 1-5 are all valid for $T > T_x$ but is invalid for $D_r \sim \tau_{r6}^{-1}$. The $D_r \sim \tau_{rn}^{-1}$ for $n$ = 1-6 are all breakdown when $T < T_x$ and in a factional form with $\xi_{5n} \simeq 0.6$. The $D_r \sim T/\eta a^3$ tested by $D_r\alpha^3 \sim \left(T\eta^2\right)^{\xi_6}$ is also established when $T$ > $T_x$ and otherwise breakdown. The differences between $\xi_1$ and $\xi_3$ imply the $a$ is temperature dependent, although the dependence is small. However, the result given by $D_r \sim T/\eta$ is almost corrected to the exact result $\xi_6 = 1.0$ when $T > T_x$ and the breakdown of $D_r \sim T/\eta$ is also reduced after taking the variation of *a* into account. It implies the importance of considering the variation of *a* in testing the SE relation and SED relation. Combining the results given by SE relation and SED relation, our simulations indicate the translational motion is still coupling with the rotational motion for $T > T_x$ and otherwise decouple. The results are due to the steric hindrance introduced by the rigid tetrahedral structure and dense packing, which makes the rotation more strongly hindered than the translation. Overall, the central finding of this study is the identification of an inverse hierarchy of dynamic arrest in a rigid tetrahedral model: contrary to OTP and most existing glass formers where translation decouples first, rotation decouples earlier than translation in our system. This demonstrates that molecular geometry can fundamentally alter the competition between translational and rotational degrees of freedom in supercooled liquids. However, there still exists some questions need to answer in the future; as we have discussed above, the volume fraction and steric hindrance have a strong influence on the decoupling order, whether there exists a critical volume fraction at which the earlier decoupling of translational dynamics relative to rotational dynamics will be reversed, like in supercooled water or OTP, especially for spherical particle system.

**Acknowledgments**

This work was supported by the National Natural Science Foundation of China (No. 12104502) and the Fundamental Research Funds for the Central Universities (No. 25CAFUC09019).

†E-mail: rengan@alumi.itp.ac.cn

## References

[1] Debenedetti P G and Stillinger F H 2001 Supercooled liquids and the glass transition *Nature* 410 259

[2] Berthier L and Biroli G 2011 Theoretical perspective on the glass transition and amorphous materials *Rev. Mod. Phys.* 83 587

[3] Ediger M D 2000 Spatially Heterogeneous Dynamics in Supercooled Liquids *Annu. Rev. Phys. Chem.* 51 99

[4] Fujara F, Geil B, Sillescu H and Fleischer G 1992 Translational and rotational diffusion in supercooled orthoterphenyl close to the glass transition *Zeitschrift für Physik B Condensed Matter* 88 195

[5] Mapes M K, Swallen S F and Ediger M D 2006 Self-Diffusion of Supercooled o-Terphenyl near the Glass Transition

Temperature *J. Phys. Chem. B* 110 507
[6] Cicerone M T, Blackburn F R and Ediger M D 1995 How do molecules move near Tg? Molecular rotation of six probes in o - terphenyl across 14 decades in time *J. Chem. Phys.* 102 471
[7] Cicerone M T and Ediger M D 1996 Enhanced translation of probe molecules in supercooled o - terphenyl: Signature of spatially heterogeneous dynamics? *J. Chem. Phys.* 104 7210
[8] Eastwood M P, Chitra T, Jumper J M, Palmo K, Pan A C and Shaw D E 2013 Rotational Relaxation in ortho-Terphenyl: Using Atomistic Simulations to Bridge Theory and Experiment *J. Phys. Chem. B* 117 12898
[9] Lombardo T G, Debenedetti P G and Stillinger F H 2006 Computational probes of molecular motion in the Lewis-Wahnström model for ortho-terphenyl *J. Chem. Phys.* 125 174507
[10] Stanley H E, Barbosa M C, Mossa S, Netz P A, Sciortino F, Starr F W and Yamada M 2002 Statistical physics and liquid water at negative pressures *Physica A: Statistical Mechanics and its Applications* 315 281
[11] Netz P A, Starr F W, Barbosa M C and Stanley H E 2002 Relation between structural and dynamical anomalies in supercooled water *Physica A: Statistical Mechanics and its Applications* 314 470
[12] Netz P A, Starr F, Barbosa M C and Stanley H E 2002 Translational and rotational diffusion in stretched water *J. Mol. Liq.* 101 159
[13] Shi R, Russo J and Tanaka H 2018 Origin of the emergent fragile-to-strong transition in supercooled water *Proc. Natl. Acad. Sci. U. S. A.* 115 9444
[14] Netz P A, Buldyrev S V, Barbosa M C and Stanley H E 2006 Thermodynamic and dynamic anomalies for dumbbell molecules interacting with a repulsive ramplike potential *Phys. Rev. E* 73 061504
[15] Debenedetti P G 2003 Supercooled and glassy water *J. Phys.: Condens. Matter* 15 R1669
[16] Shi R, Russo J and Tanaka H 2018 Common microscopic structural origin for water's thermodynamic and dynamic anomalies *J. Chem. Phys.* 149 224502
[17] Mazza M G, Giovambattista N, Stanley H E and Starr F W 2007 Connection of translational and rotational dynamical heterogeneities with the breakdown of the Stokes-Einstein and Stokes-Einstein-Debye relations in water *Phys. Rev. E* 76 031203
[18] Kawasaki T and Kim K 2019 Spurious violation of the Stokes–Einstein–Debye relation in supercooled water *Scientific Reports* 9 8118
[19] Becker S R, Poole P H and Starr F W 2006 Fractional Stokes-Einstein and Debye-Stokes-Einstein Relations in a Network-Forming Liquid *Phys. Rev. Lett.* 97 055901
[20] Edmond K V, Elsesser M T, Hunter G L, Pine D J and Weeks E R 2012 Decoupling of rotational and translational diffusion in supercooled colloidal fluids *Proc. Natl. Acad. Sci. U. S. A.* 109 17891
[21] Vivek S and Weeks E R 2017 Decoupling of translational and rotational diffusion in quasi-2D colloidal fluids *J. Chem. Phys.* 147 134502
[22] Chong S-H and Kob W 2009 Coupling and Decoupling between Translational and Rotational Dynamics in a Supercooled Molecular Liquid *Phys. Rev. Lett.* 102 025702
[23] Chang I and Sillescu H 1997 Heterogeneity at the Glass Transition: Translational and Rotational Self-Diffusion *J. Phys. Chem. B* 101 8794
[24] Berthier L, Biroli G, Bouchaud J-P, Cipelletti L and van Saarloos W 2011 *Dynamical heterogeneities in glasses, colloids, and granular media* (Oxford: Oxford University Press)
[25] Zheng Z, Wang F and Han Y 2011 Glass Transitions in Quasi-Two-Dimensional Suspensions of Colloidal Ellipsoids *Phys. Rev. Lett.* 107 065702

[26] Zheng Z, Ni R, Wang F, Dijkstra M, Wang Y and Han Y 2014 Structural signatures of dynamic heterogeneities in monolayers of colloidal ellipsoids *Nat. Commun.* 5 3829
[27] Chun D J, Oh Y and Sung B J 2021 Translation-rotation decoupling of tracers reflects medium-range crystalline order in two-dimensional colloid glasses *Phys. Rev. E* 104 054615
[28] Shi Z, Debenedetti P G and Stillinger F H 2013 Relaxation processes in liquids: Variations on a theme by Stokes and Einstein *J. Chem. Phys.* 138 12A526
[29] Lewis L J and Wahnström G 1994 Molecular-dynamics study of supercooled ortho-terphenyl *Phys. Rev. E* 50 3865
[30] Berendsen H J C, van der Spoel D and van Drunen R 1995 GROMACS: A message-passing parallel molecular dynamics implementation *Comput. Phys. Commun.* 91 43
[31] Van Der Spoel D, Lindahl E, Hess B, Groenhof G, Mark A E and Berendsen H J 2005 GROMACS: fast, flexible, and free *J. Comput. Chem.* 26 1701
[32] Nosé S 1984 A unified formulation of the constant temperature molecular dynamics methods *J. Chem. Phys.* 81 511
[33] Hoover W G 1985 Canonical dynamics: Equilibrium phase-space distributions *Phys. Rev. A* 31 1695
[34] Binder K and Kob W 2011 *Glassy materials and disordered solids: An introduction to their statistical mechanics* (Singapore: World Scientific)
[35] Debye P 1929 *Polar Molecules* (New York:The Chemical Catalog Company))
[36] Hess B 2002 Determining the shear viscosity of model liquids from molecular dynamics simulations *J. Chem. Phys.* 116 209
[37] Kob W, Donati C, Plimpton S J, Poole P H and Glotzer S C 1997 Dynamical Heterogeneities in a Supercooled Lennard-Jones Liquid *Phys. Rev. Lett.* 79 2827
[38] Mazza M G, Giovambattista N, Starr F W and Stanley H E 2006 Relation between Rotational and Translational Dynamic Heterogeneities in Water *Phys. Rev. Lett.* 96 057803
[39] Angell C A 1995 Formation of Glasses from Liquids and Biopolymers *Science* 267 1924
[40] McCall D W, Douglass D C and Falcone D R 1969 Molecular Motion in ortho‐Terphenyl *J. Chem. Phys.* 50 3839
[41] Kawasaki T and Kim K 2017 Identifying time scales for violation/preservation of Stokes-Einstein relation in supercooled water *Sci. Adv.* 3 e1700399
[42] Ren G 2022 The effective hydrodynamic radius in the Stokes–Einstein relation is not a constant *Commun. Theor. Phys.* 74 095603
[43] Chong S-H, Moreno A J, Sciortino F and Kob W 2005 Evidence for the Weak Steric Hindrance Scenario in the Supercooled-State Reorientational Dynamics *Phys. Rev. Lett.* 94 215701
[44] Xu L, Mallamace F, Yan Z, Starr F W, Buldyrev S V and Eugene Stanley H 2009 Appearance of a fractional Stokes-Einstein relation in water and a structural interpretation of its onset *Nat. Phys.* 5 565
[45] Heyes D 2019 Translational and rotational diffusion of rod shaped molecules by molecular dynamics simulations *J. Chem. Phys.* 150 184503